\documentclass[10pt, paper=a4, UKenglish]{article}
\usepackage{graphicx}

\def\Title#1{\begin{center} {\Large #1 } \end{center}}
\def\Author#1{\begin{center}{ \sc #1} \end{center}}
\def\Address#1{\begin{center}{ \it #1} \end{center}}

\newcommand\pubblock{\rightline{\begin{tabular}{l} Proceedings of the CTD/WIT 2019\\ \pubnumber\\
         \pubdate  \end{tabular}}}

\newenvironment{Abstract}{\begin{quotation} \begin{center} 
             \large ABSTRACT \end{center}\bigskip 
      \begin{center}\begin{large}}{\end{large}\end{center} \end{quotation}}

\newenvironment{Presented}{\begin{quotation} \begin{center} 
             PRESENTED AT\end{center}\bigskip 
      \begin{center}\begin{large}}{\end{large}\end{center} \end{quotation}}

\def\beq{\begin{equation}}
\def\eeq#1{\label{#1}\end{equation}}
\def\eeqn{\end{equation}}

\def\beqa{\begin{eqnarray}}
\def\eeqa#1{\label{#1}\end{eqnarray}}
\def\eeqan{\end{eqnarray}}

\let\bar=\overbar

\def\Dslash{\not{\hbox{\kern-4pt $D$}}}
\def\dslash{\not{\hbox{\kern-2pt $\del$}}}

\def\msb{{\bar{\ssstyle M \kern -1pt S}}}

\usepackage{color}
\usepackage{lineno}
\usepackage{subfig}
\usepackage{hyperref}

\usepackage{caption}
\newcommand\pubnumber{PROC-CTD19-028}

\newcommand\pubdate{\today}

\def\affiliation{
Physics Division \\
Lawrence Berkeley National Laboratory, USA}

\newcommand{\conference}{Connecting the Dots and Workshop on Intelligent Trackers (CTD/WIT 2019)\\
Instituto de F\'isica Corpuscular (IFIC), Valencia, Spain\\ 
April 2-5, 2019}

\usepackage{fancyhdr}
\definecolor{mygrey}{RGB}{105,105,105}
\begin{document}


\large
\begin{titlepage}
\pubblock

\vfill
\Title{Identifying Merged Tracks in Dense Environments with Machine Learning}
\vfill

\Author{Patrick McCormack and Milan Ganai \\
On behalf of the ATLAS Collaboration}
\Address{\affiliation}
\vfill

\begin{Abstract}
Tracking in high density environments plays an important role in many physics analyses at the LHC. In such environments, it is possible that two nearly collinear particles contribute to the same hits as they travel through the ATLAS pixel detector and semiconductor tracker. If the two particles are sufficiently collinear, it is possible that only a single track candidate will be created, denominated a ``merged track", leading to a decrease in tracking efficiency. These proceedings show a possible new technique that uses a boosted decision tree to classify reconstructed tracks as merged. An application of this new method is the recovery of the number of reconstructed tracks in high transverse momentum three-pronged $\tau$ decays, leading to an increased $\tau$ reconstruction efficiency.  The observed mistag rate is small.
\end{Abstract}

\vfill

\begin{Presented}
\conference
\end{Presented}
\vfill
\end{titlepage}
\def\thefootnote{\fnsymbol{footnote}}
\setcounter{footnote}{0}
%

\normalsize 


\renewcommand{\thefootnote}{\arabic{footnote}}

\section{Introduction}
\label{intro}

Tracking in high density environments, particularly in high energy jets and boosted $\tau$'s, plays an important role in many physics analyses at the LHC.  In such environments, it is possible that two nearly collinear particles contribute to the same position measurements (hits) as they travel through the ATLAS pixel detector and semiconductor tracker (SCT)~\cite{PERF-2007-01, Aad:2008zz, ATLAS-TDR-19, Ahmad:2007zza}.  To form tracks from hits, the pattern recognition in ATLAS combines groups of three hits into track seeds, which are extended into track candidates by adding hits from additional layers~\cite{ATL-SOFT-PUB-2007-007, Fruhwirth:1987fm}.  An ambiguity solving procedure is performed to reach a final collection of tracks: track candidates receive a ``track score'' and candidates with scores below a certain threshold are rejected~\cite{Aaboud:2017all}.  It is relevant to high density environments that if a candidate shares hits with an accepted track, its score is reduced.  Further, if the two particles are nearly collinear, it is possible that only a single track candidate will be created (a merged track), leading to a decrease in tracking efficiency.  From the study of simulated high momentum $\tau \rightarrow$ 3$\pi^{\pm}$ $\nu_{\tau}$ decays in~\cite{ATL-PHYS-SLIDE-2018-148}, it is clear that a significant number of reconstructed tracks in these events are merged, and that there is a loss of tracking efficiency that cannot be recovered by accepting more track candidates.  For example, at a $\tau~p_{\mathrm{T}}$ of 800 GeV about 10\% of events have a merged track, while a distinct track is accepted for all three pions in only about 65\% of events.  These proceedings detail a new technique: using a Boosted Decision Tree (BDT) to classify reconstructed tracks as merged~\cite{myplots}.  The following plots quantify the performance of the BDT and its application on reconstructed events.  The post-reconstruction application of the technique is also compared to other reconstruction schemes.

\begin{figure}[htb]
\centering
\subfloat[]{\includegraphics[trim={0 0 0 1cm},clip,width = 0.35\textwidth]{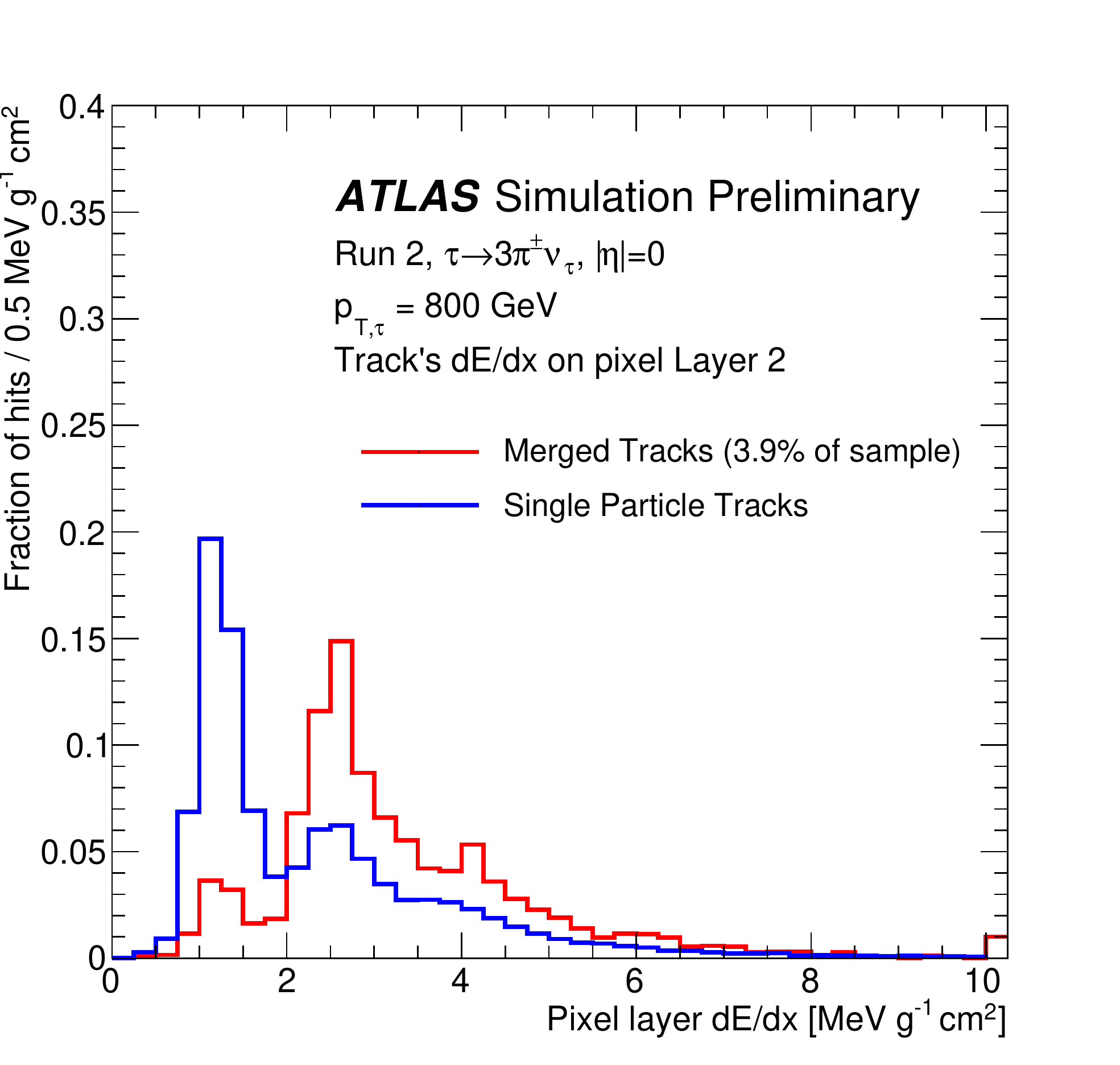}}
\subfloat[]{\includegraphics[trim={0 0 0 1cm},clip,width = 0.35\textwidth]{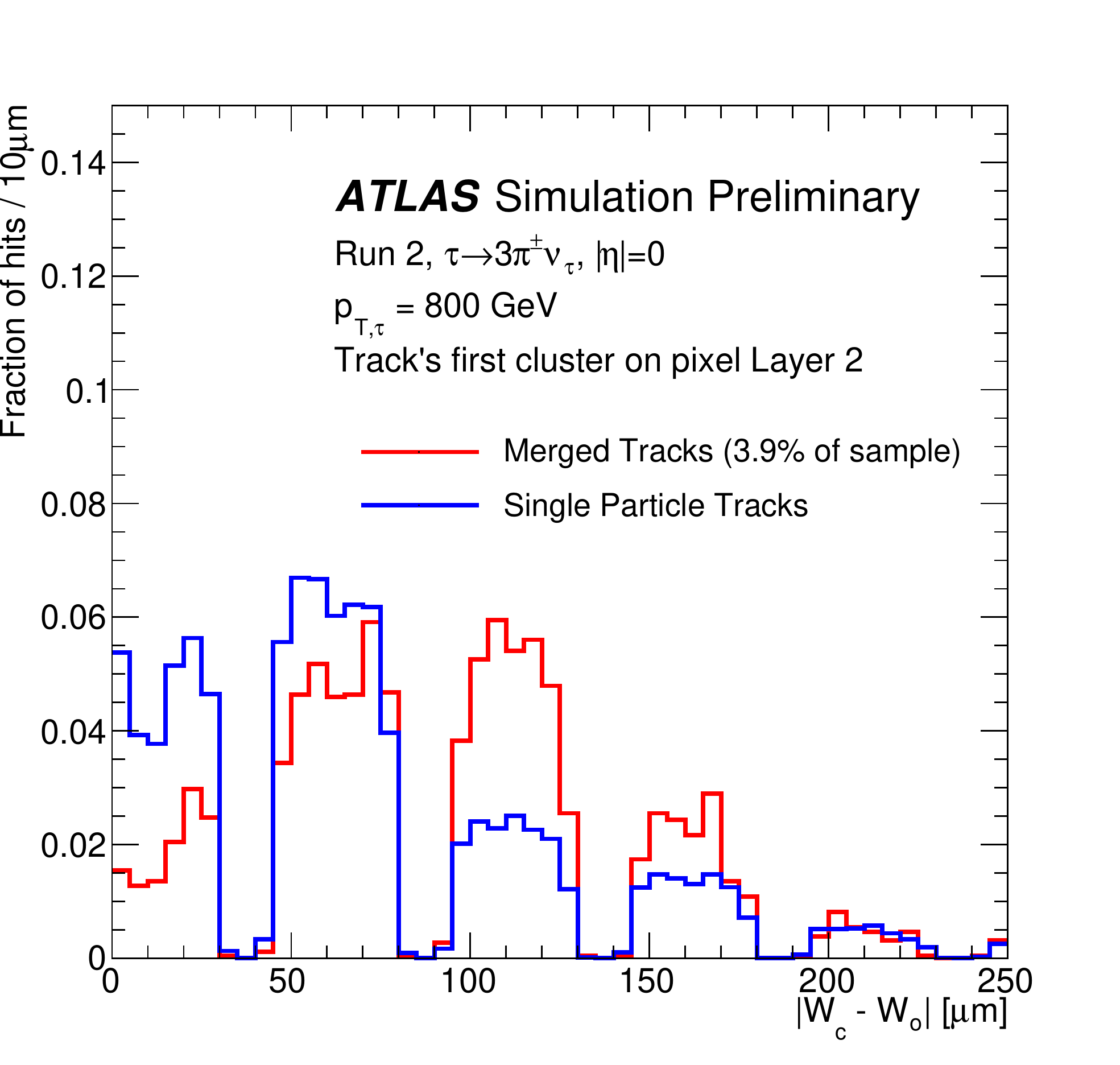}}
\subfloat[]{\includegraphics[trim={0 0 0 1cm},clip,width = 0.35\textwidth]{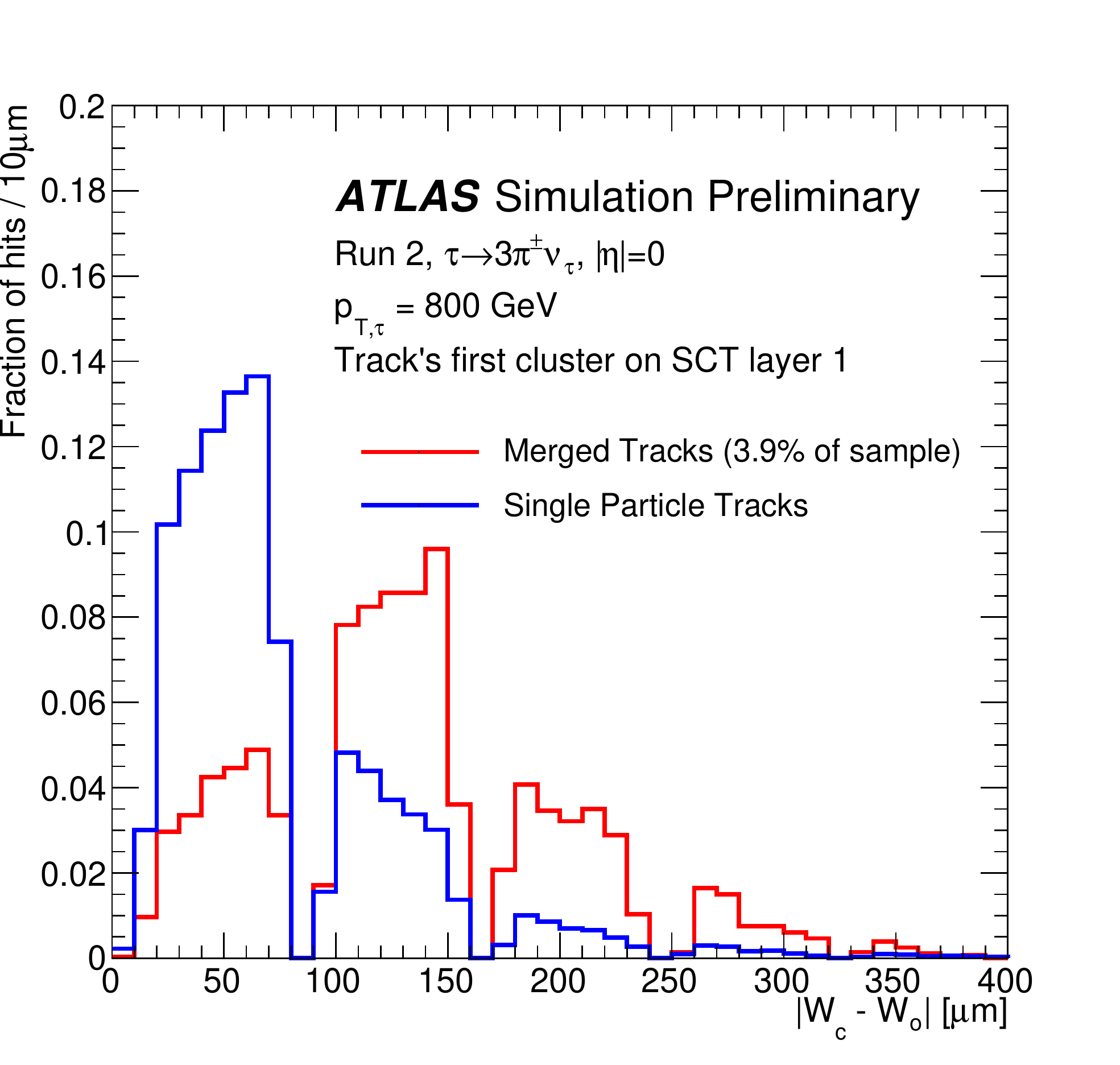}}
\caption{Illustration of the separation power of select variables that are used in the BDT.}
\centering
\label{variables}
\end{figure}

\section{Building the Boosted Decision Tree}
\label{BDT}

Because this technique is applied to fully reconstructed tracks, it can access a complete array of track variables: global track parameters, cluster information for the pixel and SCT hits along the track, and local track parameters at the hits.  The BDT uses 43 variables, shown in Table~\ref{thetable}.

Figure~\ref{variables} illustrates the separation power of a select few of the 43 variables for merged and single particle tracks.  Figure~\ref{variables}a shows the normalized distributions of the $dE/dx$ in the last layer of the pixel detector (called ``Layer 2''\footnote{The pixel detector has four layers: the ``Insertable B-Layer", the ``B-Layer", ``Layer 1", and ``Layer 2", listed from the innermost to the outermost}).  Figure~\ref{variables}b shows the normalized distributions of $|W_{\mathrm{c}} - W_{\mathrm{o}}|$, which is the difference in the expected cluster width based on track incidence angles at the detector element and the actual cluster width, for the cluster with the highest recorded charge on pixel Layer 2.  Figure~\ref{variables}c shows the normalized distributions of $|W_{\mathrm{c}} - W_{\mathrm{o}}|$ for the recorded cluster on the outer side of the innermost layer of the SCT with the lowest pull on the track parameters.  The tracks considered were reconstructed in samples of single $\tau \rightarrow$ 3$\pi^{\pm}$ $\nu_{\tau}$ events that were generated at $\eta = 0$.  Four samples were used for training and testing, with $\tau~p_{\mathrm{T}}$'s of 50, 400, 800, and 1000 GeV, each containing the same number of events.  The variable values shown here are from $\tau$'s with $p_{\mathrm{T}}$ of 800 GeV.  The variables considered show good discrimination power between the clusters of merged and single particle tracks.

\begin{table}
\small
\caption{Description of features for a BDT to identify merged tracks.  Note that there are four pixel layer and four double-sided SCT layers.}\label{tab:yourlabel}
\begin{tabular}{p{6cm}|p{1.4cm}|p{8.3cm}}
\textbf{Variable}& \textbf{Num. of Features} & \textbf{Explanation}\\
\hline
Track $p_{\mathrm{T}}$, $\eta$, $\phi$ & 3 & --\\
\hline
Num. clusters on each pixel layer & 4 & -- \\
\hline
Highest charge deposited in a cluster on each pixel layer & 4 & Sensor overlap may lead to 2 charge deposits on a single pixel layer. \\
\hline
$dE/dx$ in each pixel layer & 4 & $dE/dx$ of a track through a layer.\\ 
\hline
Boolean for whether a hit on each pixel layer is flagged as split & 4 & Determined by the pixel cluster neural network~\cite{pixneuralnet}\\
\hline
$|W_{\mathrm{c}} - W_{\mathrm{o}}|_{\mathrm{pix}}$ for pixel cluster with the highest charge on each pixel layer & 4 & $W_{\mathrm{c}}$ is the expected cluster width (the r-$\phi$ pitch) calculated from the incidence angle of the reconstructed track at the cluster in the plane perpendicular to the beam-line, the thickness of the module, and the Lorentz drift angle; $W_{\mathrm{o}}$ is the observed cluster width in integral multiples of $50~\mu$m, which is the pixel pitch.  See~\cite{ATL-PHYS-SLIDE-2018-148} for more details.\\
\hline
$|L_{\mathrm{c}} - L_{\mathrm{o}}|_{\mathrm{pix}}$ for pixel cluster with the highest charge on each pixel layer & 4 & $L_{\mathrm{c}}$ is the expected length in z-direction, and $L_{\mathrm{c}}$ is the observed length in z-direction.  Similar to~\cite{ATL-PHYS-SLIDE-2018-148}.\\
\hline
Num. clusters on each SCT layer & 4 & -- \\
\hline
$|W_{\mathrm{c}} - W_{\mathrm{o}}|_{\mathrm{SCT}}$ for SCT cluster with the lowest pull on each SCT layer & 8 & $W_{\mathrm{o}}$ is the observed cluster width in integral multiples of $80~\mu$m, which is the SCT strip pitch.  $W_{\mathrm{c}}$ is the expected cluster width: see~\cite{ATL-PHYS-SLIDE-2018-148} for more details.\\
\hline
Num. shared clusters on each SCT layer & 4 & A ``shared'' cluster is one used by multiple reconstructed tracks. \\
\end{tabular}
\label{thetable}
\end{table}

\section{Training the BDT}
\label{training}

\begin{figure}[htb]
\centering
\subfloat[]{\includegraphics[trim={0 0 0 1cm},clip,width = 0.35\textwidth]{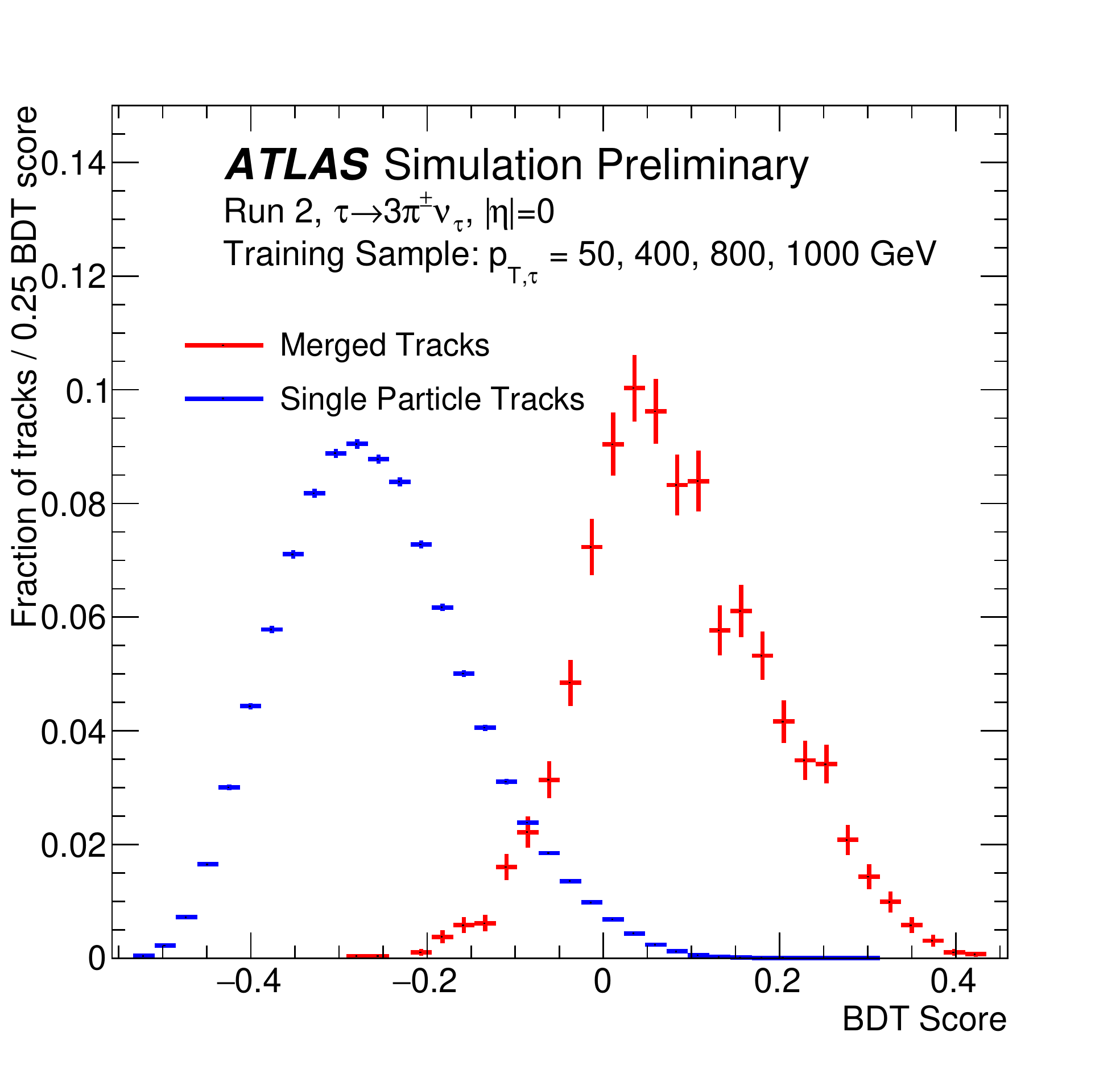}}
\subfloat[]{\includegraphics[trim={0 0 0 1cm},clip,width = 0.35\textwidth]{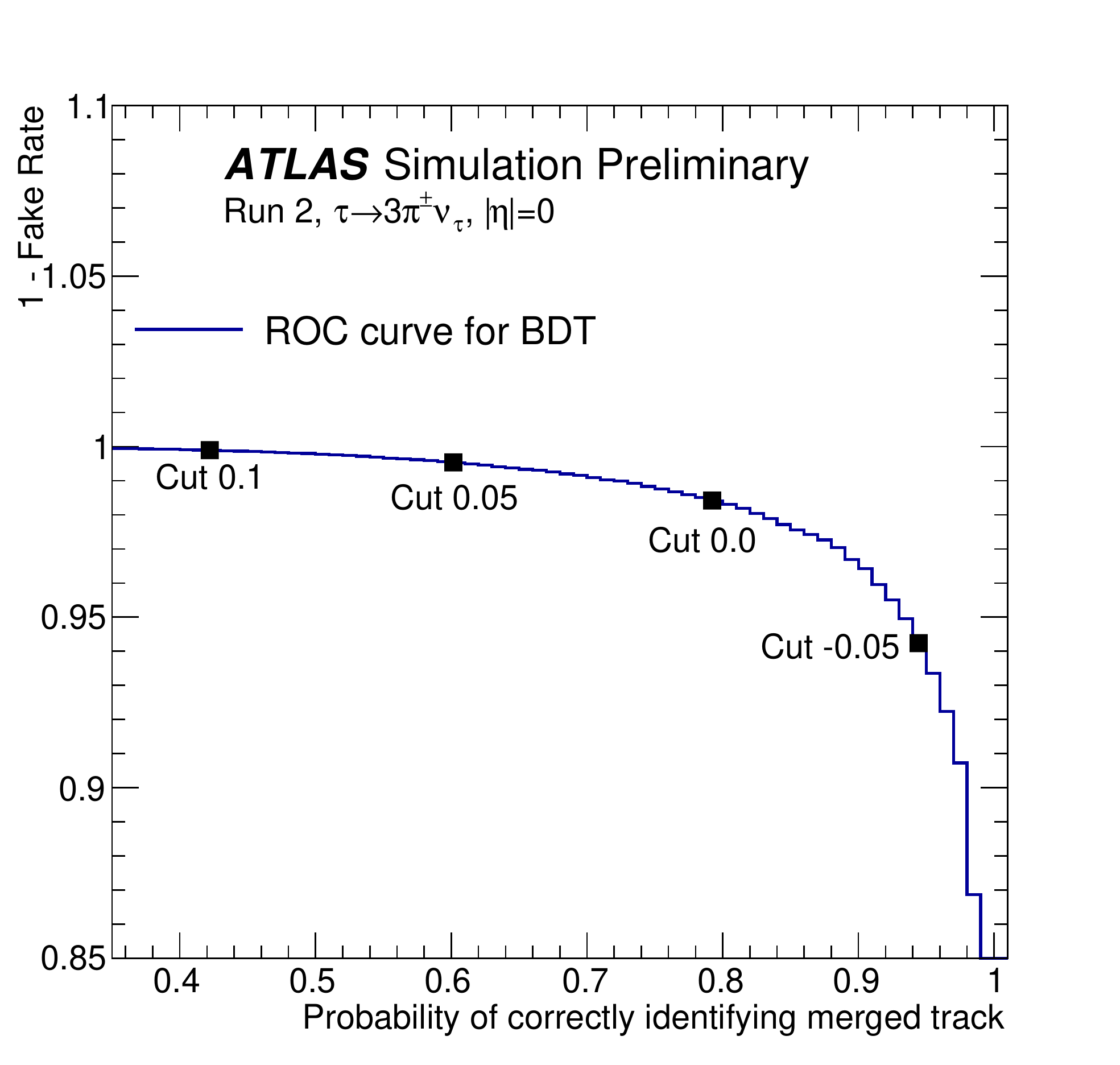}}
\caption{(a) Normalized BDT score distributions for the training sample; (b) ROC curve with points highlighting different potential cuts on BDT score.}
\centering
\label{ROC}
\end{figure}

Figure~\ref{ROC} shows the normalized BDT\footnote{The BDT uses theTMVA v.4.2.1~\cite{TMVA} implementation defaults without any hyper-parameter optimization.} score distributions for the training sample (Figure~\ref{ROC}a) and a receiver operating characteristic curve (ROC curve) with points highlighting different potential cuts on BDT score (Figure~\ref{ROC}b), where any track with a BDT score above the cut value is flagged as ``merged''.  The tracks considered come from samples of $\tau \rightarrow$ 3$\pi^{\pm}$ $\nu_{\tau}$ decays described in Section~\ref{BDT}.  In TMVA~\cite{TMVA}, signal is given a training value of 1 and background -1.  From Figure~\ref{ROC}a, it is clear that the BDT finds a significant difference between merged and single particle tracks; the error bars represent simulation statistics only.  True merged tracks are rare, representing about 3\% of tracks in the samples used.  Because of this, tracks selected with an aggressive cut on the BDT score will be predominantly single particle tracks misidentified as merged.

\section{Results: Impact on track counting}
\label{Results}

\begin{figure}[htb]
\centering
\subfloat[]{\includegraphics[trim={0 0 0 1cm},clip,width = 0.35\textwidth]{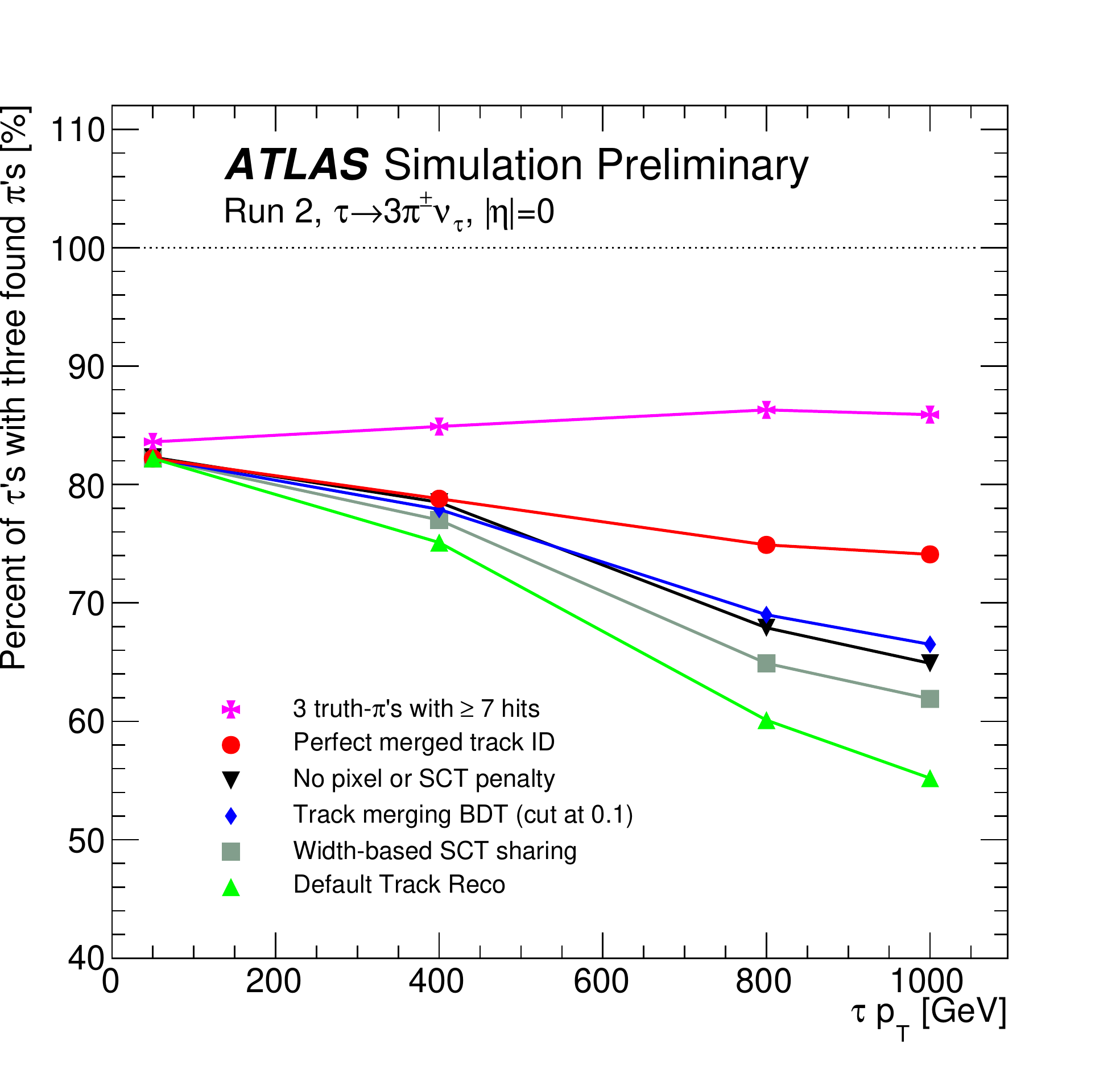}}
\subfloat[]{\includegraphics[trim={0 0 0 1cm},clip,width = 0.35\textwidth]{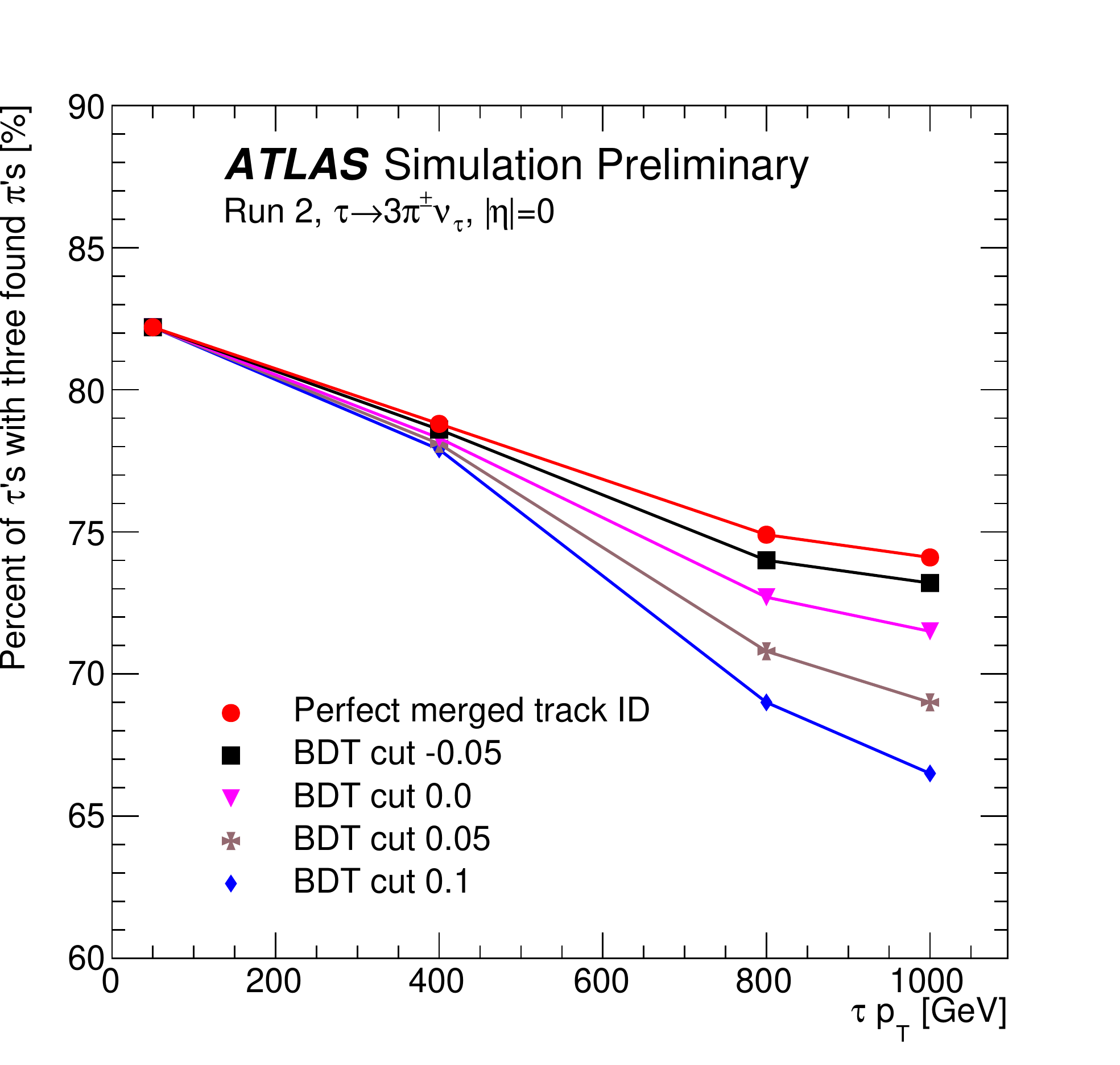}}
\subfloat[]{\includegraphics[trim={0 0 0 1cm},clip,width = 0.35\textwidth]{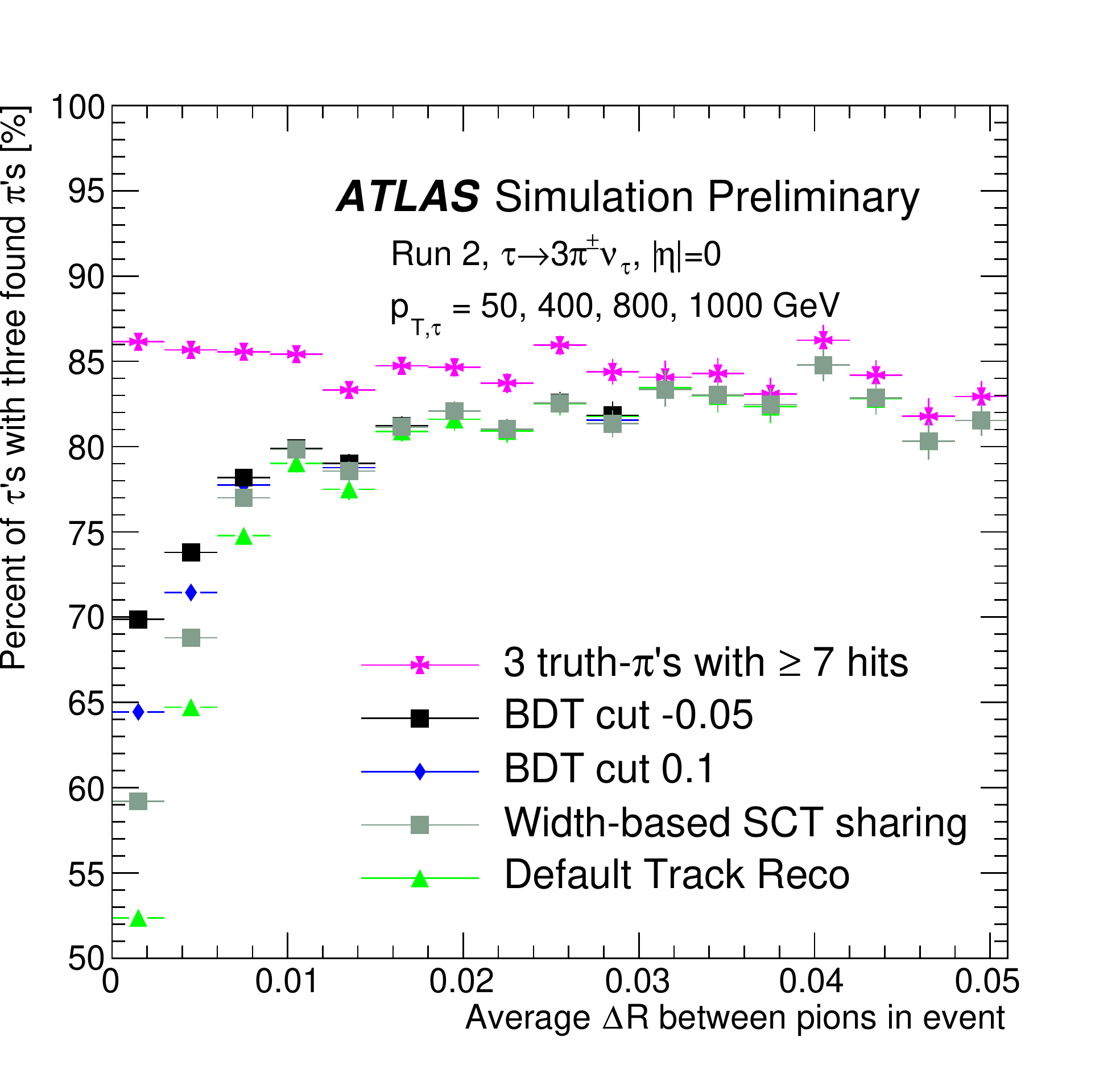}}
\caption{(a) Simulated efficiency of $\tau$ reconstruction efficiency as a function of $\tau~p_{\mathrm{T}}$ using different reconstruction configurations (a) and different BDT cuts (b). (c) shows the efficiency as a function of track density.}
\centering
\label{eff}
\end{figure}

After training the BDT, we examine it's potential impact on the tracking efficiency.  The BDT score is calculated for every track in our sample, and a track is flagged as merged if it has a BDT score above a selected cut value.  A $\tau$ event is considered fully reconstructed if all three pions from its decay are reconstructed as tracks in the event.  Both of the pions contributing to a merged track are considered to be reconstructed if that track is flagged as merged by the BDT.  Figure~\ref{eff}, shows how applying the BDT affects the efficiency for reconstructing $\tau$ events as a function of $\tau~p_{\mathrm{T}}$, both for different BDT cut values (Figure~\ref{eff}b) and as a function of track density in the event (Figure~\ref{eff}c).

In Figure~\ref{eff}a, the pink crosses show the technical reconstruction efficiency, where a pion will be considered ``found'' if it leaves at least 7 truth hits in the silicon layers of the inner detector.  The green triangles show the ATLAS default track reconstruction performance.  The pale squares show reconstruction performance using the algorithm described in~\cite{ATL-PHYS-SLIDE-2018-148}.  The inverted triangles show the reconstruction performance when the cluster sharing penalty (see Section~\ref{intro}) is turned off.  The blue diamonds show the performance when reconstructed tracks are considered to be two pions if they are flagged as merged by the BDT with a cut of 0.1, as described in Section~\ref{training}.  The BDT is applied after reconstruction.  The red circles show the performance if every truth-level merged track is counted as two pions.  This filter is also applied after track reconstruction.

\section{Mistag rate}
\label{Fakes}

\begin{figure}[htb]
\centering
\subfloat[]{\includegraphics[trim={0 0 0 1cm},clip,width = 0.35\textwidth]{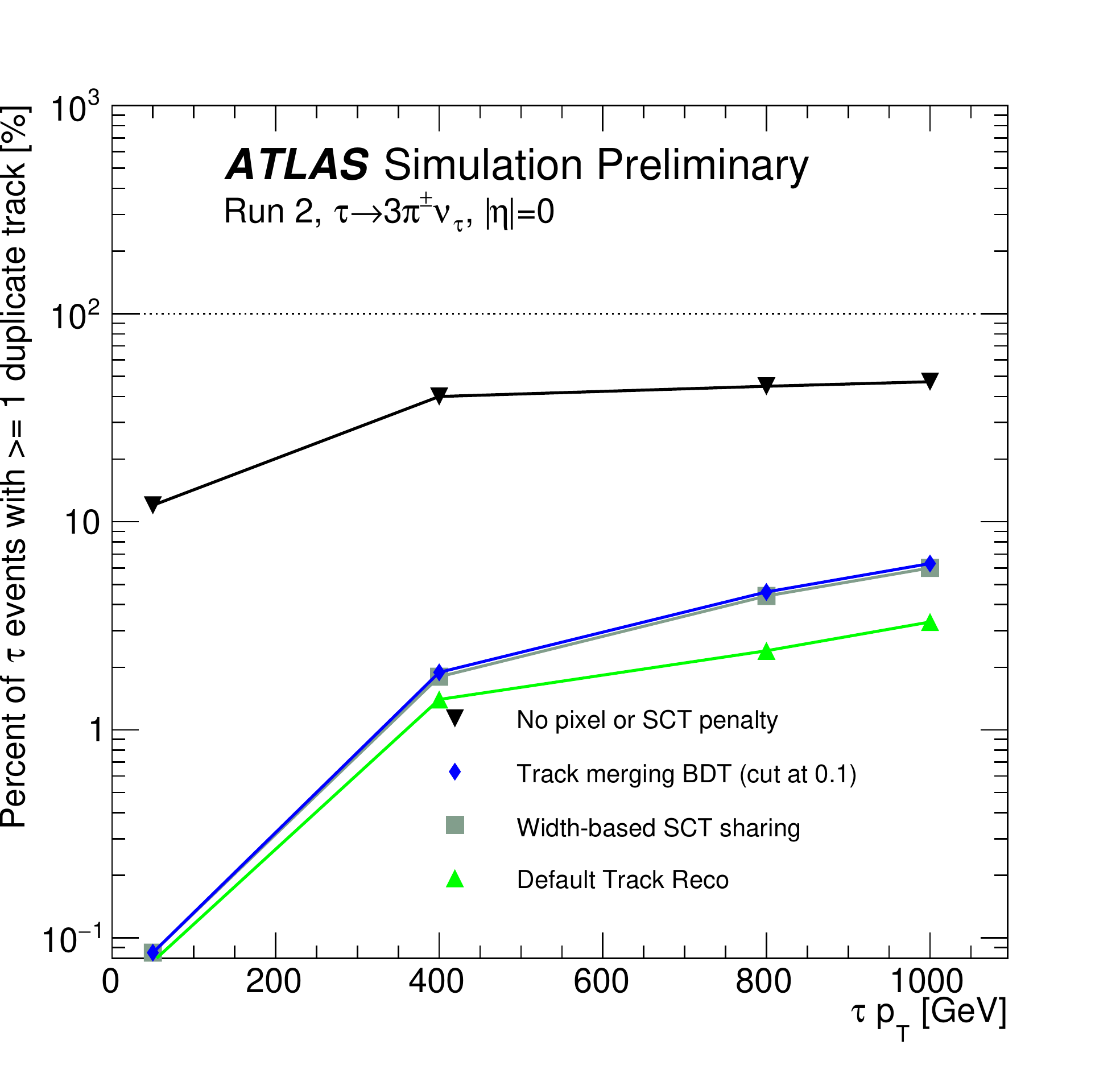}}
\subfloat[]{\includegraphics[trim={0 0 0 1cm},clip,width = 0.35\textwidth]{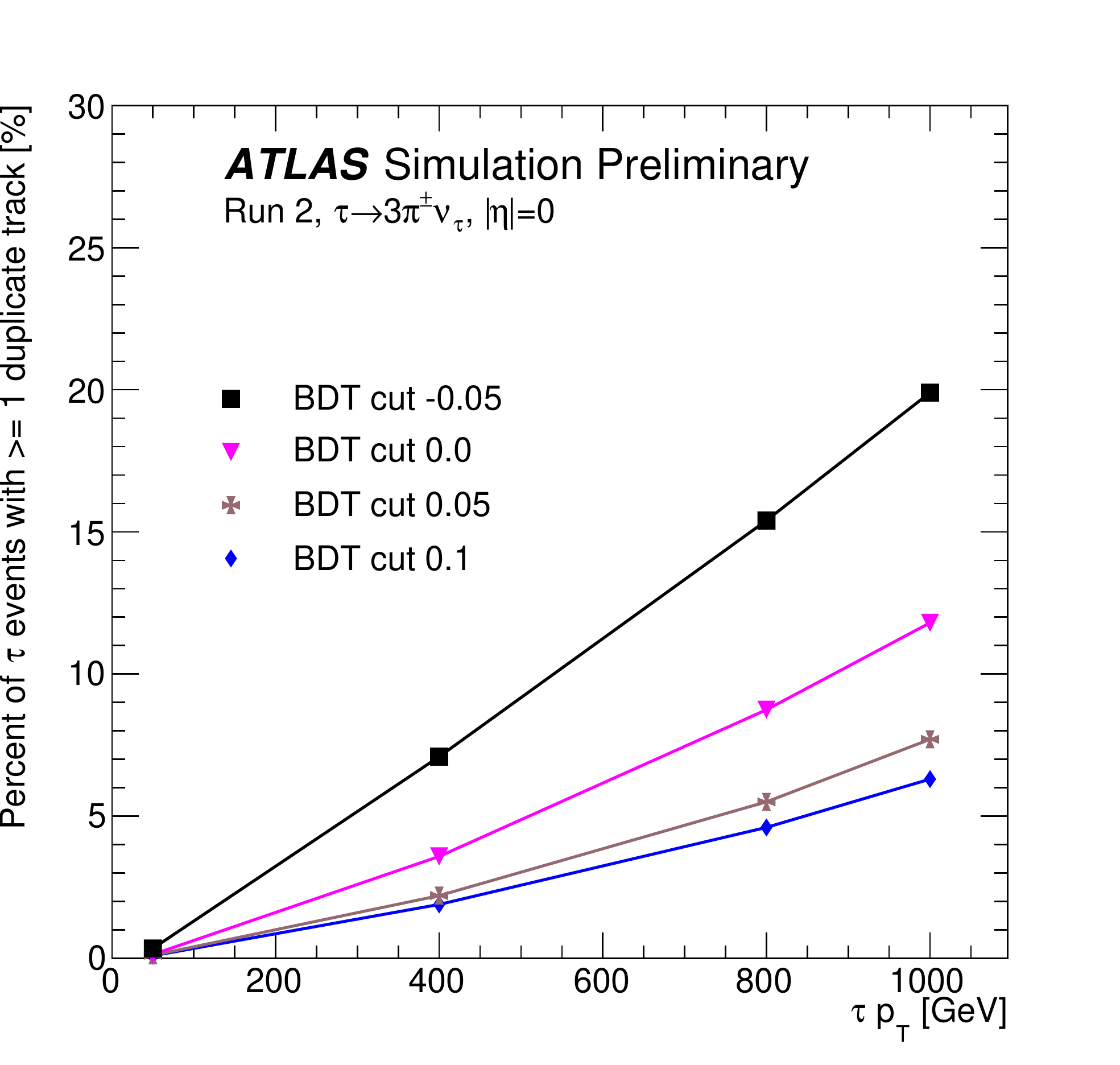}}
\caption{(a) Simulated rate of accepting a ``duplicate'' track as a function of $\tau~p_{\mathrm{T}}$ using different reconstruction configurations (a) and different BDT cuts (b).}
\centering
\label{fakeplots}
\end{figure}

Incorrectly tagging a single-particle track as merged would distort an otherwise correct measurement.  For example, if merged tracks are counted as two particles with $p_{\mathrm{T}}$ equal to the original track, an event with an incorrectly split track could essentially have an extra high-$p_{\mathrm{T}}$ track that is unrelated to an actual particle.  Thus it is important to find a balance between increased reconstruction efficiency (as seen in Sec.~\ref{Results}) and potential mistags.  Figure~\ref{fakeplots} shows the impact implementing the BDT has on the ``duplicate rate'' for the same sample used in Figure~\ref{eff}.  An event is considered to have at least one ``duplicate'' track if at least one pion is truth-matched~\cite{truthmatch} to more than one reconstructed track.  Figure~\ref{fakeplots}a shows the impact as a function of $\tau~p_{\mathrm{T}}$, where the line and marker styles are the same as described for Figure~\ref{eff}.  Tracks reconstructed with the algorithm which allows track candidates to share SCT clusters if the cluster has an anomalous $|W_{\mathrm{c}} - W_{\mathrm{o}}|$ have a few percent increase in the duplicate rate over the default algorithm~\cite{ATL-PHYS-SLIDE-2018-148}.  The BDT was trained and tested on tracks reconstructed using that SCT cluster-sharing algorithm, so the ``Track merging BDT'' line should be compared to the ``Width-based SCT sharing'' line.  Applying the BDT with a cut of 0.1 creates a negligible increase in the duplicate rate.  Figure~\ref{fakeplots}b illustrates the duplicate rate when using different BDT cuts.

\begin{figure}[htb]
\centering
\subfloat[]{\includegraphics[trim={0 0 0 1cm},clip,width = 0.35\textwidth]{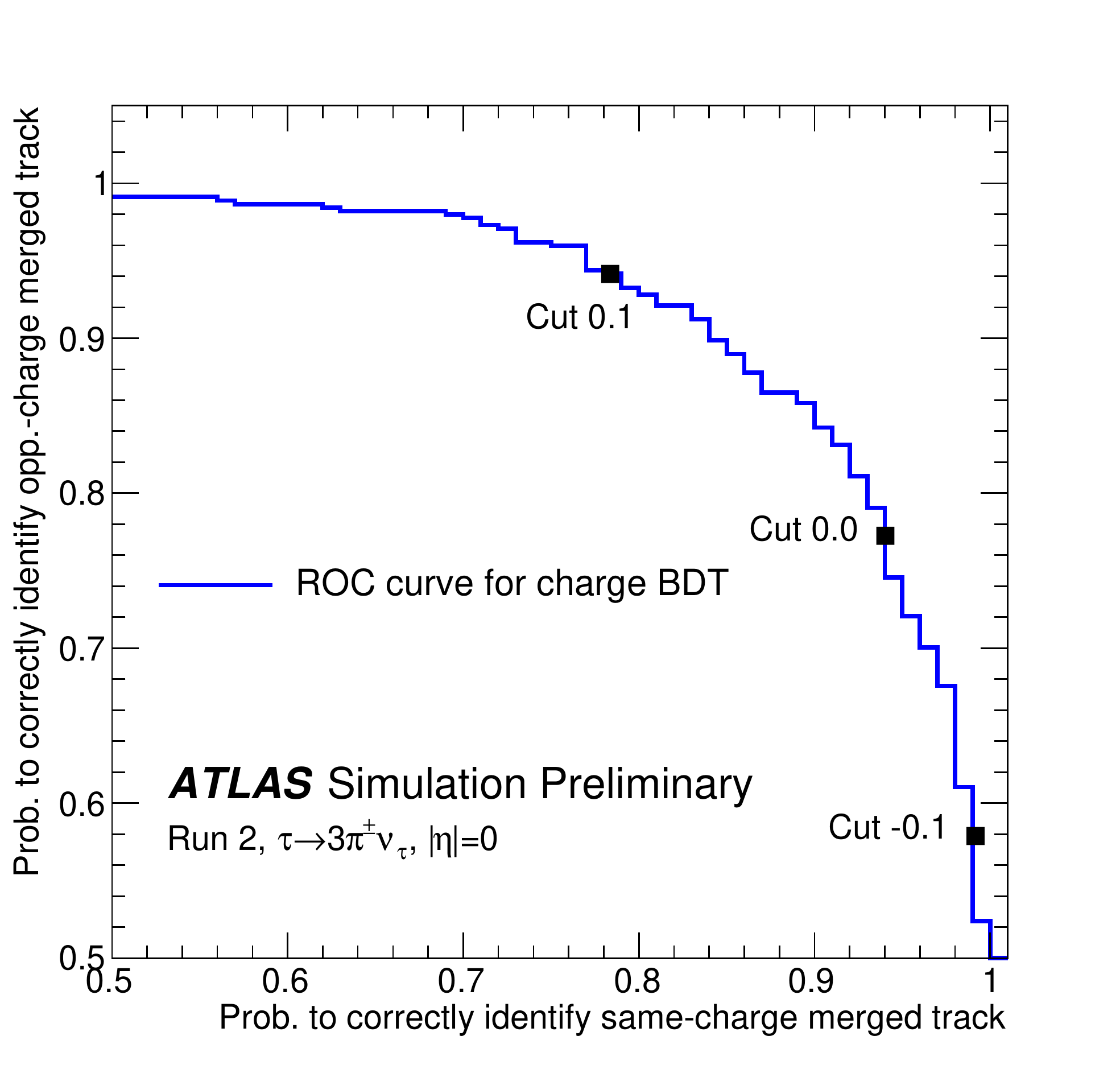}}
\subfloat[]{\includegraphics[trim={0 0 0 1cm},clip,width = 0.35\textwidth]{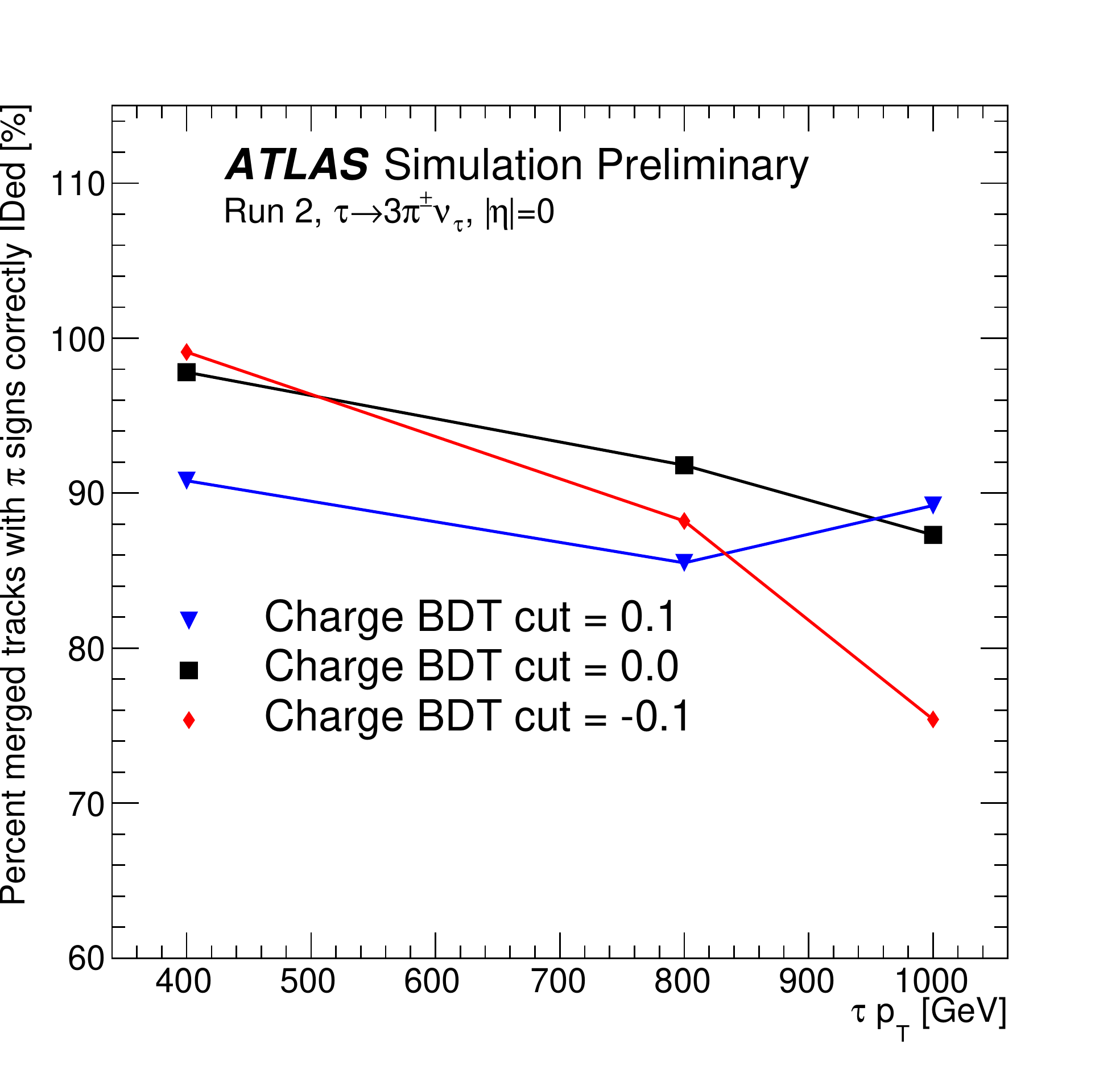}}
\caption{(a) ROC curve for the charge BDT; (b) efficiency for correctly identifying charges in a merged track as a function of $\tau~p_{\mathrm{T}}$.}
\centering
\label{chargeplots}
\end{figure}

\section{Identifying the particle charges}
\label{ChargeID}

A natural question to ask is whether the variables fed into the BDT contain enough information to perform a more sophisticated analysis than a simple merged/unmerged classifier.  For example, we have created a BDT to distinguish whether the pions in a merged track have the same or opposite electric charge.  This secondary BDT uses the same 43 variables as in Table~\ref{thetable} and is trained on tracks which are flagged as merged by the initial BDT and considered merged at truth level.  Figure~\ref{chargeplots} demonstrates that it is possible to correctly identify the pion charges with over 85\% accuracy for a wide range of $\tau$ $p_{\mathrm{T}}$; \ref{chargeplots}a shows a ROC curve for correctly identifying a merged track as coming from same-sign or opposite-sign $\pi$'s, and \ref{chargeplots}b shows the charge-tagging efficiency as a function of $\tau~p_{\mathrm{T}}$ for different charge BDT settings.  A cut of 0 is the most efficient over a widest range of $p_{\mathrm{T}}$'s.

\section{Conclusions and Outlook}
\label{Conclusions}

These proceedings have demonstrated a possible new technique for finding tracks in the ATLAS detector that are actually created by two nearly collinear particles.  Using a BDT, one can find about 50\% of merged tracks while only increasing the duplicate rate at the sub-percent level.  Moving forward, this technique can be investigated in other environments where collinearity is expected, such as high-$p_{\mathrm{T}}$ jets; making statements about the impact on measureables such as $\tau$ tagging or jet $p_{\mathrm{T}}$ resolution will require further study.

\section{Acknowledgements}
\label{acknowledgements}

This work was supported by the U.S. Department of Energy, Office of Science under contract DE-AC02-05CH11231.  Patrick McCormack was supported by the National Science Foundation Graduate Research Fellowship under Grant No. DGE 1752814. Any opinion, findings, and conclusions or recommendations expressed in this material are those of the author and do not necessarily reflect the views of the National Science Foundation.


\begin{thebibliography}{99}


  
\bibitem{PERF-2007-01}
  ATLAS Collaboration,
  ``The ATLAS Experiment at the CERN Large Hadron Collider,"
  JINST {\bf 3}, S08003 (2008).
  
\bibitem{Aad:2008zz}
  Aad, G. et al.,
  ``ATLAS pixel detector electronics and sensors,"
  JINST {\bf 3}, P07007 (2008).
  
\bibitem{ATLAS-TDR-19}
  ATLAS Collaboration,
  ``ATLAS Insertable B-Layer Technical Design Report,"
  ATLAS-TDR-19, https://cds.cern.ch/record/1291633 (2010).
  
\bibitem{Ahmad:2007zza}
  Ahmad, A. et al.,
  ``The Silicon microstrip sensors of the ATLAS semiconductor tracker,"
  Nucl. Instrum. Meth. {\bf A578}, 98-118 (2007).
  
\bibitem{Abat:2008zza}
  Abat, E. et al.,
  ``The ATLAS Transition Radiation Tracker (TRT) proportional drift tube: Design and performance,"
  JINST {\bf 3}, P02013 (2008).
  
\bibitem{ATL-SOFT-PUB-2007-007}
  ATLAS Collaboration,
  ``Concepts, Design and Implementation of the ATLAS New Tracking (NEWT),"
  ATL-SOFT-PUB-2007-007, http://cdsweb.cern.ch/record/1020106 (2007).
  
\bibitem{Fruhwirth:1987fm}
  Fruhwirth, R.,
  ``Application of Kalman filtering to track and vertex fitting,"
  Nucl. Instrum. Meth. {\bf A262}, 444-450 (1987).
  
\bibitem{Aaboud:2017all}
  ATLAS Collaboration,
  ``Performance of the ATLAS Track Reconstruction Algorithms in Dense Environments in LHC Run 2,"
  Eur. Phys. J. {\bf C77}, 673 (2017)
  [arXiv:1704.07983 [hep-ex]].
  
\bibitem{ATL-PHYS-SLIDE-2018-148}
  ATLAS Collaboration,
  ``Splitting Strip Detector Clusters in Dense Environments,"
  ATL-PHYS-SLIDE-2018-148, https://cds.cern.ch/record/2311048 (2018).
  
\bibitem{pixneuralnet}
  ATLAS Collaboration,
  ``A neural network clustering algorithm for the ATLAS silicon pixel detector,"
  JINST {\bf 9}, P09009 (2014)
  [arXiv:1406.7690 [hep-ex]].
  
\bibitem{TMVA}
  A. Hoecker et al.,
  ``TMVA - Toolkit for Multivariate Data Analysis,"
  CERN-OPEN-2007-007, arXiv:physics/0703039 (2009).
  
\bibitem{myplots}
  ATLAS Collaboration,
  ``Identifying merged tracks with machine learning,"
  IDTR-2019-003, https://atlas.web.cern.ch/Atlas/GROUPS/PHYSICS/PLOTS/IDTR-2019-003/ (2019).
  
\bibitem{truthmatch}
  ATLAS Collaboration,
  ``Early Inner Detector Tracking Performance in the 2015 data at $\sqrt{s}$ = 13 TeV,"
  ATL-PHYS-PUB-2015-051, http://cdsweb.cern.ch/record/2110140 (2015).
  
\end{thebibliography}
\end{document}